\documentclass[epj]{svjour}
\usepackage{graphics}
\newcommand{\ket}[1]{|#1\rangle}
\begin{document}
\title{Simultaneous position and state measurement of Rydberg atoms}

\author{C. S. E. van Ditzhuijzen\inst{1} \and A. F. Koenderink\inst{2} \and L. D. Noordam\inst{1} \and H. B. van Linden van den Heuvell\inst{1}
}                     
%
%
\institute{Van der Waals-Zeeman Institute, University of
Amsterdam, Valckenierstraat 65, 1018 XE Amsterdam, The Netherlands
\and Center for Nanophotonics, FOM Institute AMOLF, Kruislaan 407,
1098 SJ Amsterdam, The Netherlands}
\date{Published online 21 June 2006 – Eur. Phys. J. D \textbf{40}, 13-17. DOI: 10.1140/epjd/e2006-00140-1}
%
\abstract{ We present a technique for state-selective position
detection of cold Rydberg atoms. Ground state Rb atoms in a
magneto-optical trap are excited to a Rydberg state and are
subsequently ionized with a tailored electric field pulse. This
pulse selectively ionizes only atoms in e.g. the 54d state and not
in the 53d state. The released electrons are detected after a slow
flight towards a micro channel plate. From the time of flight of
the electrons the position of the atoms is deduced. The state
selectivity is about 20:1 when comparing 54d with 53d and the
one-dimensional position resolution ranges from 6 to 40 $\mu$m
over a range of 300 $\mu$m. This state selectivity and position
resolution are sufficient to allow for the observation of coherent
quantum excitation transport.
\PACS{
      {32.80.Pj, 34.60.+z, 32.80.Rm}{}
     } 
} 

\authorrunning{C. van Ditzhuijzen et al.}
\titlerunning{Simultaneous position and state measurement of Rydberg atoms}
\maketitle
\section{Introduction}
\label{intro}

Rydberg atoms are well-known for their large dipole moments.
Strong dipole-dipole interaction leads to interesting novel
phenomena in particular for cold atoms. Examples are the
triggering of spontaneous plasma formation \cite{plasma,li} and
the so called dipole blockade of optical excitation of a Rydberg
atom in the vicinity of an already present Rydberg atom
\cite{jaksch,lukin,eyler,weidemuller}. Another example is the
formation of macrodimers from two Rydberg atoms
\cite{macrodimers,mol_res}.

When the distances between the interacting Rydberg atoms are
fixed, which is achieved by using laser cooled atoms, the
dipole-dipole interaction can be coherent
\cite{robicheaux,pillet}. For a pair of atoms with two successive
principal quantum numbers, both in the highest (or lowest) energy
Stark state, the interaction leads to an exchange of states
between the atoms; in other words an excitation hops from one atom
to the other \cite{robicheaux}. The time required for hopping of
an excitation back and forth is given by:
\begin{equation}
t_{hop}=9 \pi \frac{d^3}{n^4}
\end{equation}
with $t_{hop}$ in atomic units, $d$ the distance between the atoms
in atomic units and $n$ the principal quantum number. For $n$=60
and $d$=20 $\mu$m the hopping time is 2.8 $\mu$s. These mesoscopic
numbers make a quantum information system with position-resolved
Rydberg atoms feasible. An interesting system \cite{robicheaux} is
for example the hopping of one excitation (e.g. n=61 or $\ket{1}$
in qubit notation) on a lattice of lower Rydberg states (e.g. n=60
or $\ket{0}$). Another example is the diffusion of the $\ket{1}$
state in a random gas of atoms in state $\ket{0}$, which might
show features of Anderson localization \cite{anderson}.

In this paper we present experimental results that are a first
step towards exploration of the coherent evolution of systems of
Rydberg atoms. We show that the position-dependent readout of
$\ket{0}$ and $\ket{1}$ states is indeed possible. The possibility
to create and measure Rydberg atoms in a position and state
resolved way is demonstrated here by exciting a narrow region in a
cold cloud of rubidium atoms to the 54d state. The one-dimensional
position sensitive detection is done with a time-of-flight
technique. Time of flight is our choice in preference to a
position-sensitive multi-channel plate detector because of
simplicity and compatibility with existing equipment.

The time-of-flight approach makes it difficult to combine these
measurements with other time-resolved techniques. State resolution
is therefore based on the threshold character of field ionization,
rather than timing as in conventional ramped state selective field
ionization (SFI) \cite{gallagher}. The combination of techniques
leads to conflicting optimizations. In the remainder of this paper
these problems will be quantified and solved.

\section{State and position determination}
\label{sfitof}

We first focus on the state determination by SFI \cite{gallagher}.
The electric field $E$ in which a Rydberg atom ionizes is roughly
given by the classical ionization threshold, which is at an energy
of $U=-2\sqrt{E}$, in atomic units, and corresponds to a field of
$E=1/16n^4$. Due to Stark shifts, however, the energy of the state
changes with the electric field. Therefore the ionization field is
different. In non-hydrogenic atoms many Stark states couple with
each other, forming a level scheme with many avoided crossings.
The best state selectivity is achieved when the crossings are
traversed either all adiabatically or all diabatically. In the
latter case the field has to be ramped up faster than technically
convenient, so we choose to ramp the field very slowly. In the
adiabatic case the energy stays approximately the same and the
ionization field will just be around $1/16n^4$.

\begin{figure}
\resizebox{\columnwidth}{!}{
  \includegraphics{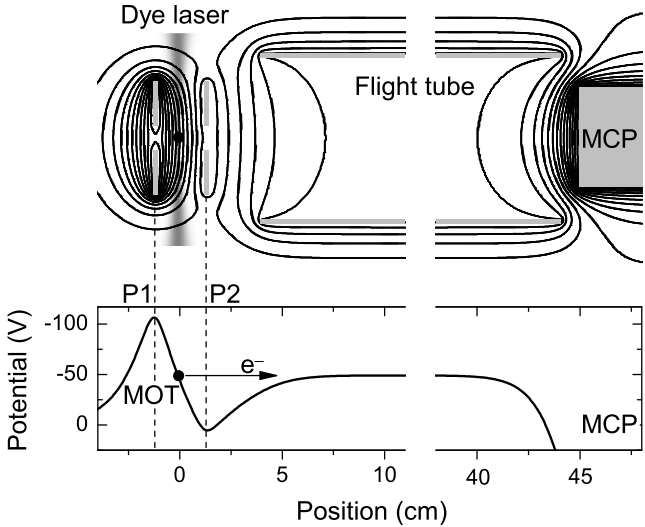}
} \caption{Schematic representation of the components that provide
the electric field together with equipotential lines, which are
calculated numerically. The applied voltages are: -115 V on the
first plate (P1), +10 V on the second plate (P2), -48 V on the
flight tube and +90 V on the mesh in front of the MCP. Below is a
graph of the electric potential on axis. Also depicted are the
position of the MOT (black dot) and the dye laser beam.}
\label{veldlijn}
\end{figure}

For the time-of-flight method we use the same electric field that
ionizes the atom. This electric field pushes the electron through
one of the ionizing electrodes into a field-free flight tube. The
time it takes for the electron to fly through the tube is a
measure for its kinetic energy at the beginning of the tube, and
therefore a measure for the potential energy at its starting
position. At the position of the atom cloud the electric potential
decreases linearly with position (see Fig. \ref{veldlijn}).
Consequently the electric potential, and thus the arrival time of
the electron on the detector is a measure for the original
position of the Rydberg atom. To magnify the arrival time
differences beyond the instrumental time resolution we use a long
flight tube and take care that the electrons are slow during
flight. This is achieved by setting a potential on the flight tube
that is almost the same as the potential at the starting position.
The arrival time of the electron as a function of the initial
position $x$ is approximately
\begin{equation}\label{eq:toftime}
t(x)=L\sqrt{\frac{m_e}{2e(E x+V_0)}}
\end{equation}
with $L$ the length of the flight tube $e$ the charge and $m_e$
the mass of the electron, $E$ the electric field and $V_0$ the
potential at $x$=0 minus the potential on the flight tube.

For the time-of-flight technique the time of ionization has to be
well defined. This is in contradiction with the slow field ramp
needed for the state selective field ionization. Therefore we use
a more advanced scheme for the combination of the two techniques:
a slow field ramp brings the field just below the ionization limit
of the $\ket{1}$ state, followed by a fast pulse to go above the
limit of the $\ket{1}$ state. In addition, this pulse has to stay
below the ionization limit of the $\ket{0}$ state, in order to
detect the atoms in state $\ket{1}$ exclusively.

\section{Experimental Setup}
\label{setup}

To make sure the atoms don't move on the relevant timescale of the
experiment we use a magneto-optical trap (MOT) of $^{85}$Rb atoms
as our cold atom source. A typical rubidium MOT has a temperature
below 300 $\mu$K \cite{mot}, which corresponds to an average
velocity of 0.3 $\mu$m$/\mu$s. The atoms are loaded from a
dispenser into the MOT, created at the intersection of three
orthogonal pairs of counterpropagating $\sigma^+$ - $\sigma^-$
laser beams at the center of a magnetic quadrupole field. The
background pressure is 3 10$^{-8}$ mbar. Charged particles from
the dispenser are removed by deflection in an electric field of 50
V/cm, which is shielded from its environment. The laser frequency
is tuned about 13 MHz below the 5S$_{1/2}$ (F=3) $\to$ 5P$_{3/2}$
(F=4) resonance. A repumping laser beam tuned to the 5S$_{1/2}$
(F=2) $\to$ 5P$_{1/2}$ (F=3) resonance is added at the MOT center.
The cooling laser and the repumping laser are both Toptica DL100
diode lasers at resp. 780 nm and 795 nm and are locked to the
rubidium resonance with Doppler-free FM spectroscopy. During all
measurements the cooling laser is blocked, so that all atoms are
pumped to the 5S$_{1/2}$ (F=3) ground state by the repumping
laser.

To create atoms in a Rydberg state we excite cold atoms with an 8
ns, $\sim$2 $\mu$J laser pulse at 594 nm by a two-photon process
from the 5s state to 53d or 54d (resp. $\ket{0}$ and $\ket{1}$).
This light is provided by a Lambda Physik dye laser pumped with a
Spectra Physics frequency-doubled, Q-switched Nd:YAG laser with a
repetition rate of 10 Hz. The linewidth of the dye laser pulse is
0.15 cm$^{-1}$, which is well below the energy spacing between 53d
an 54d ($1.55$ cm$^{-1}$). A small contamination of the nearest
s-state, which lies 0.32 cm$^{-1}$ higher, can not be excluded but
is of no further relevance. The dye laser beam is focused in the
MOT cloud by a lens placed on a micrometer translation stage. The
beam waist is determined to be approximately 23(1) $\mu$m ($1/e$
diameter), measured with the knife-edge technique. The
polarization of the light is parallel to the electric field that
will be applied after the laser pulse.

The Rydberg atoms are field ionized by applying voltage pulses on
two field plates P1 and P2 spaced by 2.5 cm (see Fig.
\ref{veldlijn}). The circular stainless steel plates have a
diameter of 5.5 cm and have a 1.4 cm hole in the middle for the
transmission of one of the pairs of counterpropagating MOT laser
beams. The released electrons go through the positively charged
field plate P2 into a 40 cm long stainless steel tube, which is at
negative potential and serves to slow down the electrons. After a
45 cm flight the electrons are detected on a Hamamatsu Micro
Channel Plate (MCP). A copper mesh is placed in front of the MCP.
The mesh is 95\% open and set at +90 Volt.

Additional coils outside the vacuum chamber compensate for
background magnetic field. This is important for the electrons to
fly straight towards the detector. The axis of the magnetic
quadrupole field for the MOT is in the same direction as the
electric field between the field plates.

\begin{figure}
\resizebox{\columnwidth}{!}{
  \includegraphics{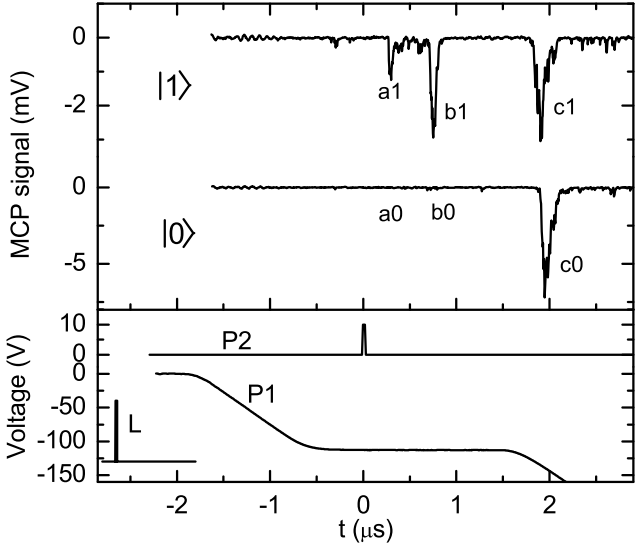}
} \caption{In the lower panel the applied voltage pulses on the
plates P1 and P2 are shown. The laser pulse is shown as L. In the
upper panel the MCP signal for the $\ket{0}$ and $\ket{1}$ state
(resp. 53d and 54d) are depicted. The first two peaks a and b are
both resulting from the fast pulse on P2 at t=0. The peaks c
originate from the atoms that are left over. The voltage on the
flight tube is -48 V and the position of the dye laser beam is 280
$\mu$m on the scale of Fig. \ref{tof}. } \label{sfiplaatje}
\end{figure}

For the simultaneous position and state detection we use the pulse
scheme depicted in the lower half of Fig. \ref{sfiplaatje}.
Shortly after the dye laser pulse we apply a slow voltage ramp on
the field plate P1 which goes from 0 V to -115 V in 1.2 $\mu$s. It
then remains -115 V for 1.3 $\mu$s, while a 10 V pulse is applied
on the other plate P2 with a duration of 30 ns and a rise time of
10 ns. After that the voltage on the plate P1 is increased to -230
V in 1.2 $\mu$s, to ionize all atoms that remained unionized. With
the fast pulse we aim to ionize the upper state atoms exclusively
at a well defined time. The slow ramp is provided by an Agilent
33240A arbitrary waveform generator amplified by a home-built
amplifier system. The 10 V pulse is provided by an HP 8114A high
power pulse generator.

For the calibration of the conversion from time-of-flight to
position we shift the Rydberg-production volume along the flight
path by moving the lens that focuses the dye laser beam in the MOT
cloud and record the signal of the MCP for every lens position.
The voltage on the flight tube is carefully tuned, such that the
electrons created at the edge of the MOT nearest to the plate P2
are still clearly detected on the MCP. This usually means that
they have a flight time of ~1 $\mu$s, or equivalently, an energy
of ~0.5 eV. At lower velocities the signal density drops.

Before describing our experimental results, we consider the effect
of the timing of our experiment on the achievable position
resolution. First the thermal spreading of atoms during the slow
ramp of the electric field is approximately 0.9 $\mu$m, assuming a
MOT temperature of 300 $\mu$K. A second concern is that atoms may
move during the ramp due to electrostatic forces on their dipole
moments \cite{merkt}. These forces can arise from the field
gradient associated with small inhomogeneities in the applied
electric field. In our setup the gradient is approximately 20
V/cm$^2$ at a field of 50 V/cm, which gives a displacement of less
than 0.06 $\mu$m over the duration of the slow field ramp. Both
numbers are negligible compared to the required position
resolution of 20 $\mu$m.

\section{Results}
\label{results}

In the upper panel of Fig. \ref{sfiplaatje} the MCP signal during
the voltage pulses is depicted for two different dye laser
wavelengths. The upper trace is the signal for $\lambda$=594.166
nm, which excites to the 54d state or the $\ket{1}$ state and the
trace below this is the signal for $\lambda$=594.193 nm, leading
to excitation of the 53d state or the $\ket{0}$ state.

\begin{figure}
\resizebox{\columnwidth}{!}{
  \includegraphics{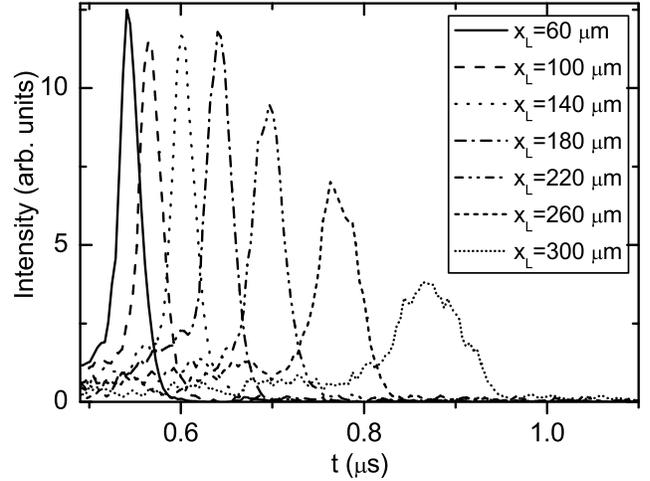}
} \caption{Measured time-of-flight spectra for different laser
focus positions $x_L$. The peaks correspond, apart from a minus,
to the peak b1 as defined in Fig. \ref{sfiplaatje}. The time t=0
is when the 30 ns pulse is applied to plate P2.} \label{tof}
\end{figure}

{In the upper trace of Fig. \ref{sfiplaatje} three different peaks
can be distinguished, labelled a, b and c. Apparently atoms,
originally in the 54d state, can be in different states at the time
of ionization. This is because the atoms have traversed the avoided
crossings in the Stark map differently during the slow field ramp,
ending up in a distribution over blue and red states in the Stark
manifold. The fast pulse sequentially ionizes some of these states
within a very short time window (during the 10 ns rise time of the
pulse). As the potential at the time of electron release is
different, electrons acquire different velocities depending on
whether they originate from red or blue states, which results in
peak a (the fast electrons released first) and peak b (the slowest
electrons). Peak c results from atoms in a red state that remain
unionized by the fast pulse and are ionized by the subsequent ramp.
This slow ramp is used for estimating the ionization efficiency
during the earlier pulses.}

For the time-of-flight measurements we will focus on peak b,
because this peak is most sensitive to the position of the
Rydberg-production volume. Peak b1 consists of 30\% of the total
signal of state $\ket{1}$. When the atoms are prepared in state
$\ket{0}$, the same measurement should ideally give no signal
(b0). From our measurements we find an upper bound that is 1\% or
2\% of the total signal. The ratio of the genuine and spurious
$\ket{1}$ signal is therefore about 20. In future experiments we
will attempt to improve this ratio further. The spurious signal
could result from atoms that are actually excited to the $\ket{1}$
state, due to the finite laser linewidth, or from atoms in the
$\ket{0}$ state that ionize in a smaller field.

Fig. \ref{tof} shows time-of-flight spectra for the state
$\ket{1}$. The peaks correspond to the peak b1 as defined in Fig.
\ref{sfiplaatje} for several positions of the dye laser focus. It
can be seen that for faster electrons the peaks are narrower, but
also start to overlap more. The peaks in the time-of-flight
spectrum appeared to be narrowest when the duration of the fast
pulse applied on plate P2 was 30 ns.

\begin{figure}
\resizebox{\columnwidth}{!}{
  \includegraphics{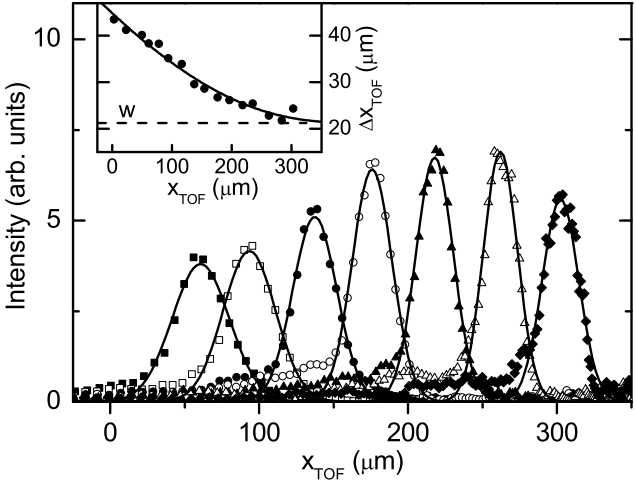}
} \caption{The peaks of Fig. \ref{tof} are transformed to the
position domain using Eq. \ref{eq:toftime}. Solid lines show
gaussian fits. In the inset the $1/e$-widths of these gaussians
$\Delta x_{TOF}$ are plotted against the center position
$x_{TOF}$. The solid line is a fit to the data with Eq.
\ref{eq:widthfit}. The dashed line $w$ shows the fitted width of
the Rydberg-production volume (21 $\mu$m).} \label{tofx}
\end{figure}

We transformed the time-of-flight spectra to the position domain
using Eq. \ref{eq:toftime} with $L$=40 cm. As the potential is
significantly more complicated (see Fig. \ref{veldlijn}) than
assumed for the simple model of Eq. \ref{eq:toftime}, we use an
effective field $E$ and $V_0$ as fit parameters. Also we add a
short offset time of 13 ns, based on numerical calculations, for
the parts of the flight that are outside the flight tube. Our fit
is based on more traces than depicted in Fig. \ref{tof}, since the
measurements were taken for lens positions 20 $\mu$m apart. The
peaks in the position domain can be excellently fitted with
gaussian profiles, as shown in Fig. \ref{tofx}. Optimal agreement
of the center positions of the fitted gaussians with the known 20
micron between lens positions is obtained for $E$=50.2 V/cm and
$V_0$=-2.2 V.

It can be seen that in the position domain the peaks get narrower
for larger $x$ (i.e. "downhill" on the potential in Fig.
\ref{veldlijn}), while the corresponding peaks in the time domain
(i.e. later times) get wider. For large $x$ the broadening due to
the finite width of the Rydberg-production volume $w$ (due to a
finite laser beam waist) is dominant. For small $x$ the time
broadening effect $\tau$ is dominant. This time width $\tau$ and
not the Rydberg-production volume $w$ is the limit for the spatial
resolution of our time-of-flight system. Therefore we decouple
$\tau$ from the 2$\sigma$-widths $\Delta x_{TOF}$ of the gaussians
by fitting the two broadening effects to the following expression
\begin{equation}\label{eq:widthfit}
\Delta x_{TOF} = \sqrt{w^2+\left(\frac{dx(t)}{dt}\right)^2
\tau^2}
\end{equation}
with $x(t)$ the inverse function of Eq. \ref{eq:toftime}. In the
inset of Fig. \ref{tofx} $\Delta x_{TOF}$ is plotted against the
center position $x_{TOF}$ together with the fitted dependence. The
fitted parameters are $\tau$=20(1) ns and $w$=21(1) $\mu$m, which
is comparable to $1/\sqrt{2}$ times the measured laser waist (the
$1/\sqrt{2}$ factor originates from the two-photon character
excitation).

\begin{figure}
\resizebox{\columnwidth}{!}{
  \includegraphics{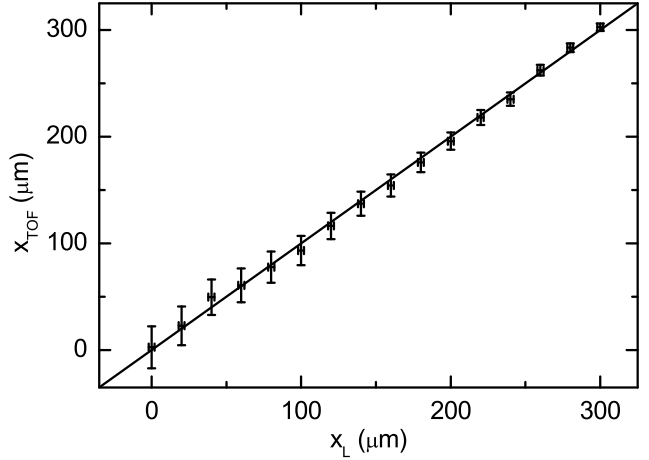}
} \caption{The center positions of the fitted gaussians of Fig.
\ref{tofx} $x_{TOF}$ plotted against the read-off position $x_L$.
The solid line is $x_{TOF}=x_L$. The small horizontal error bar is
an estimate of the read-off error (2 $\mu$m). The vertical error
bars are obtained by transforming the fitted time width $\tau$ to
position and give the spatial resolution of our system.}
\label{xtoferror}
\end{figure}

In Fig. \ref{xtoferror} we plot the center positions of the
gaussians $x_{TOF}$ against the read-off position of the lens that
focuses the dye laser beam into the MOT cloud ($x_L$), together
with the line $x_{TOF}=x_L$. The data points show an excellent
agreement with the straight line, which shows that the simple
model Eq. \ref{eq:toftime} is well suited to convert flight times
into position. The small horizontal error bar is an estimate of
the read-off error (2 $\mu$m). The vertical error bars provide the
spatial resolution of our system and are obtained by transforming
the time width $\tau$ to position. The spatial resolution ranges
from 6 to 40 $\mu$m over a distance of 300 $\mu$m ($1/e$
diameter), and it is better than 20 $\mu$m over a range of 150
$\mu$m.

\section{Conclusions}
\label{conc}

We have demonstrated a technique for state-selective position
detection of cold Rydberg atoms. The state selectivity is about
20:1 when comparing 54d with 53d and might be improved by using a
narrower spectral linewidth of the Rydberg-exciting laser or by
optimizing the electric field pulse sequence. The position
resolution ranges from 6 to 40 $\mu$m over a distance of 300
$\mu$m. These n-state and position resolution are sufficient to
allow for the observation of long range hopping transport by
coherent dipole-dipole interactions in cold Rydberg systems
\cite{robicheaux}.

\bigskip

\small{We thank F. Robicheaux for fruitful discussions and R.
Kemper, A. de Snaijer and A. G\"urtler for help in building the
experimental setup. We thank the FOM Institute AMOLF for the loan
of equipment. This work is part of the research program of the
"Stichting voor Fundamenteel Onderzoek der Materie" (FOM), which
is financially supported by the "Nederlandse Organisatie voor
Wetenschappelijk Onderzoek" (NWO).}

\end{document}